\documentclass[11pt]{article}
\usepackage{graphicx}
\usepackage{amsmath,amssymb}
\newcommand{\comment}[1]{{}}

\title{String theory in a vertex operator representation:
a simple model for testing loop quantum gravity}

\author{Artem Starodubtsev\thanks{email: astarodu@astro.uwaterloo.ca} \\
 \\ \centerline{\footnotesize \it Department of Physics, University of Waterloo} \\ \centerline{\footnotesize \it 200
University ave W, Waterloo, ON, Canada N2L 3G1} \\
\centerline{\footnotesize \it and} \\
\centerline{\footnotesize \it Perimeter Institute for
Theoretical Physics} \\
\centerline{\footnotesize \it 35 King st N, Waterloo, ON, Canada
N2J 2W9 }}

\date{}

\begin{document}

\maketitle

\begin{abstract}
The loop quantum gravity technique is applied to the free bosonic
string. A Hilbert space similar to loop space in loop quantum
gravity as well as representations of diffeomorphism and
hamiltonian constraints  on it are constructed. The string in this
representation can be viewed as a set of interacting relativistic
particles each carrying a certain momentum. Two different
regularizations of the hamiltonian constraint are proposed. The
first of them is anomaly-free and give rise to interaction  very
similar to that of two dimensional $\phi^4$-model. The second
version of hamiltonian constraint is similar to $\phi^3$-model and
contains an anomaly. A possible relation of these two models to
the conventional quantization of the string based on Fock space
representation is discussed.
\end{abstract}

\section{Introduction}
It is well known that all the attempts to apply standard
perturbative quantum field theory techniques to general relativity
yield results which are inconsistent. Ultraviolet divergencies can
not be cancelled via renormalization. There are two possible ways
to treat this situation. One of them is to accept that general
relativity can not be quantized as it is and should be considered
as a low energy limit of another theory which would have better
ultraviolet behavior while quantized in the usual way. Another
possibility is to admit that the problem is not with general
relativity but with the application of standard quantization
techniques to it. One then hopes that if we developed an
appropriate quantization procedure which would take into account
the main features of general relativity such as diffeomorphism
invariance there would be no ultraviolet problem at all. The later
has been proved to be the case in the approach called loop quantum
gravity (for review see \cite{Rovelli_rev 97}), a mathematically
well defined explicitly background independent framework for the
description of quantum space time. Loop quantum gravity is defined
as a representation of the Poisson algebra of Wilson loop
observables \cite{RovelliSmolin88,RovelliSmolin90}. This algebra
is well suited to a diffeomorphism invariant theory as Wilson loop
observables can be defined without using a background metric.

The choice of the loop algebra as the basis for the quantization
makes the resulting theory drastically different from usual
quantum field theory based on the Fock representation. By
requiring that Wilson loops be well defined operators on the
Hilbert space of the theory we assume that certain states
concentrated on lower dimensional (in the present case one
dimensional) structures such as loops and graphs have finite norm.
In particular this makes such an approach not applicable to
background dependent (non diffeomorphism invariant) field theory.
The loop states in this case would span a non-separable state
space which cannot be made sense of in quantum theory. It is
diffeomorphism invariance that reduce non-separable loop space to
separable knot space.

A natural question then arises: how to relate the loop space
representation and Fock representation generally used in low
energy physics. The answer to this question is probably necessary
to understand low energy limit of loop quantum gravity which is
presumably described by perturbative field theory in background
Minkowskian space-time. However the very perturbation theory for
quantum gravity is inconsistent and need to be modified we don't
know ahead how. Therefore in the case of gravity we don't have a
consistent low energy quantum field theory at hand to relate the
loop space quantization with.

It would therefore be useful to consider a model to which both
loop space and Fock space quantization could be applied. It is not
however straightforward to find such a model. Fock space
quantization can be applied to a linear theory or to a  theory
which has a consistent perturbative expansion around the linear
approximation. Loop space quantization can be applied to a
diffeomorphism invariant theory. It is however possible in some
cases to construct Hilbert space in a background independent way
for a background dependent theory. This possibility initiated some
research on relation between loop space and Fock space for Maxwell
theory \cite{Varadarajan2000,AshtekarLewandowski2001}. However the
Fock space in Maxwell theory is background dependent and therefore
the relation between it and loop space has to be looked for at the
kinematical, non diffeomorphism invariant level. In this case one
needs to relate Fock space with non-separable loop space.

In the present paper we consider a quantization of the free
bosonic string. The string has the property that it is both linear
and diffeomorphism invariant. Fock space quantization of the
string is well developed. Contrary to background dependent
theories the physical Hilbert space of the string does not depend
on a parameterization of the string world-sheet. Therefore one can
try to relate this Fock space to an analog of loop space of the
string at the diffeomorphism invariant level. It can be shown that
a suitable generalization of loop space quantization can also be
applied to the string. This should provide a good stage on which
Fock space and loop space quantization can be compared to each
other.

This paper is organized as follows. In section 2 the basics of the
kinematics of loop quantum gravity are reviewed. The scheme is
cast in a generalized form which can be applied to a theory the
basic configuration variables of which are not necessary
connections. On the other hand all the details related to gauge
symmetry are skipped. In section 3 this scheme is applied to the
string. The kinematical Hilbert space is constructed and the
kernel of the diffeomorphism constraint operator is found. In
section 4 a regularization of the hamiltonian constraint is
suggested and the its action  on  kinematical states is defined.
In section 5 an analog of spin foam model based on the expansion
of the projector operator on the kernel of the given hamiltonian
constraint is constructed. This model turns out to be very similar
to the two-dimensional $\phi^4$ model. In section 6 an alternative
version of the hamiltonian constraint is proposed. This version
gives rise to a model with $\phi^3$ interaction.  It is argued
that this model has more chances to reproduce the ordinary string
theory based on the Fock space quantization. In section 7 we
discuss possible ways to relate the resulting picture to the
conventional string theory, as well as to string bit models
\cite{SusskindKlebanov88,Thorn93}.

\section{Canonical quantization of a diffeomorphism invariant theory}
The canonical action of a general diffeomorphism invariant theory
can be written in the form
\begin{equation}
S=\int dt \int_\Sigma [ P \wedge \dot Q - H -\lambda_\alpha C^\alpha],
\label{dia}
\end{equation}
where $Q$ is an $n$-form field defined on a spacelike
$d-1$-dimensional manifold $\Sigma$ which plays the role of
configuration variable, $P$ is a $d-n-1$-form field canonically
conjugated to $Q$, $H$ is the Hamiltonian, and $C^\alpha$ are
constraints. The canonical commutation relation between $Q$ and
$P$ can be written by using index free notation as
\begin{equation}
\{P(x),Q(y)\}=V^{D-n-1}(x) \wedge V^{n}(y) \delta(x-y),
\label{ccr}
\end{equation}
where  $V^{d-n-1}(x)$ and $V^{n}(y)$ are $d-n-1$- and $n$-
dimensional volume forms respectively and $\delta(x-y)$ is zero
weight delta function. Neither action nor canonical commutation
relation involve any background metric. Canonical variables $Q$
and $P$ may also have intrinsic indices and there may be
symmetries related to them. Intrinsic symmetries may play an
important role in further constructions. In particular in the case
of general relativity in the connection representation the
intrinsic gauge symmetry lead to the appearance of objects such as
Wilson loops and spin networks in quantum theory. However for
shortness in the present paper we will ignore the details related
to intrinsic symmetries and  consider canonical variables having
no intrinsic indices.

The general setup for canonical quantization of such a theory can be described as follows. First one can pick up an
arbitrary $n$-dimensional submanifold $\Gamma$ of $\Sigma$. The canonical coordinate $Q(x)$ can be integrated over this submanifold
without using any background metric structure, which give rise to the following integral canonical variable
\begin{equation}
Q_{\Gamma}=\int_{\Gamma} Q(x).
\label{qdef}
\end{equation}
$Q_{\Gamma}$ is invariant with respect to diffeomorphisms which
leave $\Gamma$ unchanged. Similarly, one can take a
$d-n-1$-dimensional submanifold $\Xi$ and define a variable
\begin{equation}
P_{\Xi} = \int_{\Xi} P(x).
\label{pdef}
\end{equation}
Canonical commutation relations between these integral variables have the following explicitly background independent form
\begin{equation}
\{ P_{\Xi},Q_{\Gamma} \} = int[\Xi,\Gamma],
\end{equation}
where $int[\Xi,\Gamma ]$ is the number of 0-dimensional intersections between $\Xi$ and $\Gamma$.

To define the kinematical Hilbert space of the theory one first
introduces the space $L$ of cylindrical states. The cylindrical
state $\Psi_{\{ \Gamma \},f }(Q)$ is a function of configuration
variable of the form
\begin{equation}
\Psi_{\{ \Gamma \},f }(Q)=f(U_{\Gamma_1}(Q), U_{\Gamma_2}(Q),...,U_{\Gamma_N}(Q)),
\label{cf}
\end{equation}
where $\{ \Gamma \}$ is a set of $n$-dimensional submanifolds
embedded in $\Sigma$ and $U_{\Gamma_\alpha}(Q)=
\exp(iQ_{\Gamma_\alpha})$. On the space of cylindrical functions
one can define scalar product. Two states $\Psi_{\{ \Gamma \},f} $
and $\Psi_{\{ \Gamma ' \},g} $ are orthogonal if $\{ \Gamma \}
\not= \{ \Gamma ' \} $ and
\begin{equation}
(\Psi_{\{ \Gamma \},f},\Psi_{\{ \Gamma \},g})= \int dU_1...dU_n {\overline{ f(U_1,...,U_N)}}g(U_1,...,U_N).
\label{scpr}
\end{equation}
The Hilbert completion of the space of cylindrical functions with
respect to the scalar product (\ref{scpr}) is called auxiliary
Hilbert space $H_{aux}$. It is on this space where all the
operators of the theory are defined. One can introduce an
orthonormal basis on $H_{aux}$
\begin{equation}
\Psi_{\{ \Gamma \}, \{ k \} } = \prod\limits_{\alpha} U_{\Gamma_\alpha}^{k_\alpha},
\label{basst}
\end{equation}
\begin{equation}
(\Psi_{\{ \Gamma \},\{ k \} },\Psi_{\{ \Gamma' \},\{ k' \}})=\delta_{\{\Gamma \} \{\Gamma' \}}
\delta_{\{ k \} \{ k' \}}
\end{equation}
Depending on geometry of configuration space $k$ may take either
on a discrete or on a continuous set of values.

The Hilbert space $H_{aux}$ carries a natural unitary representation ${\cal U}(Diff)$ of the diffeomorphism group of $\Sigma$
\begin{equation}
[{\cal U}(\phi) \Psi] (Q) = \Psi(\phi^{-1} Q), \ \ \ \ \ \phi \in { Diff}.
\end{equation}
A diffeomorphism transformation sends basis state (\ref{basst}) to
another basis state of the form (\ref{basst})
\begin{equation}
[{\cal U}(\phi )\Psi_{\{ \Gamma \},\{ k \}}](Q)=\Psi_{\{ \Gamma \},\{ k \}} (\phi^{-1}Q) = \Psi_{\{ \phi \Gamma \},\{ k \}}(Q)
\end{equation}
The space $H_{diff}$ of the solutions of diffeomorphism constraint
is formed by the states invariant under ${\cal U}$. Because such
states have infinite norm in $H_{aux}$ generalized functions
techniques is used in their construction. $H_{diff}$ is defined
first as a linear subset of $L^*$, the topological dual of $L$. It
is then promoted to a Hilbert space by defining a suitable scalar
product over it. $H_{diff}$ is the linear subset of $L^*$ formed
by the linear functionals $\rho$ such that
\begin{equation}
\rho({\cal U}(\phi )\Psi)=\rho(\Psi)
\label{dusp}
\end{equation}
for any $\phi \in Diff$. From now on one can adopt bra/ket notations. One can write (\ref{dusp}) as
\begin{equation}
\langle \rho \vert {\cal U}(\phi) \Psi \rangle = \langle \rho \vert \Psi \rangle
\end{equation}
and write the basis state (\ref{basst}) $\Psi_{\{ \Gamma \}, \{ k \} }$ as $\vert \{ \Gamma \}, \{ k \}\rangle$.
We denote as $\{ \gamma \} (\{ \Gamma \})$ the equivalence class of a set of submanifolds  embedded into $\Sigma$
under the action of diffeomorphism group to which $\{ \Gamma \}$ belongs. Each such class defines an element
$\langle \{ \gamma \}, \{ k \} \vert$ of $H_{diff}$ via
\begin{eqnarray}
\langle \{ \gamma \}, \{ k \} \vert \{ \Gamma \}, \{ k' \} \rangle=
\left\{
\begin{array}{ccc}
0 & if & \{ \gamma \} \not= \{ \gamma \} (\{ \Gamma \}) \\
c_\gamma \delta_{\{ k \} \{ k' \} } & if & \{ \gamma \} \not= \{ \gamma \} (\{ \Gamma \})\\
\end{array} \right.
\end{eqnarray}
Here $c_\gamma$ is the integer number of isomorphisms (including the identity) of the set of submanifolds $\{ \Gamma \}$ into
itself that can be obtained from a diffeomorphism of $\Sigma$.
A scalar product is then naturally defined in $H_{diff}$ by
\begin{equation}
\langle \{ \gamma \}, \{ k \} \vert  \{ \gamma' \}, \{ k' \} \rangle \equiv
\langle \{ \gamma \}, \{ k \} \vert  \{ \Gamma' \}, \{ k' \} \rangle
\label{discpr}
\end{equation}
for an arbitrary $\{ \Gamma \}$ such that $\{ \gamma \}( \{ \Gamma' \} ) = \{ \gamma' \}$.

\section{Kinematical Hilbert space and diffeomorphism constraint of the free bosonic string}
In this section all the constructions of the previous section are
applied to the free bosonic string.
 The action of the string can be written in a form similar to (\ref{dia}),
\begin{equation}
S=\int dt \int\limits_{0}^{2\pi} d\sigma [p_\mu \dot x^\mu -
\lambda \tilde{\tilde{D}} -N \tilde{\tilde{ {\cal H}}}].
\end{equation}
Here $x^\mu$ are embedding parameters of the string world-sheet
into target space which are scalars with respect to spacial
diffeomorphisms and $p_\mu d\sigma$ are 1-form valued conjugated
momenta,
\begin{equation}
\tilde{\tilde{D}}={\partial_\sigma x^\mu} p_\mu \label{difcon}
\end{equation}
is the constraint generating spacial diffeomorphisms and
\begin{equation}
\tilde{\tilde{{\cal H}}}=p_\mu p^\mu + T^2{\partial_\sigma x^\mu}
{\partial_\sigma x_\mu} \label{hcon}
\end{equation}
is the Hamiltonian constraint generating timelike diffeomorphisms,
$T$ being the string tension. In both cases double tilde denotes
density weight 2. Commutation relations between $x^\mu$ and $p_\mu
d\sigma$ have the following form
\begin{equation}
\{ p_\mu (\sigma) d\sigma, x^\nu (\sigma')\}= \delta_\mu^\nu \delta (\sigma, \sigma') d\sigma .
\end{equation}

One can introduce variables similar to (\ref{qdef},\ref{pdef}).
First, because $x^\mu$ are scalars (or 0-forms) with respect to
the diffeomorphisms of the string space-line they give rise to
0-dimensional objects (points $\sigma$ at string space-line) and
the corresponding canonical variables will be simply $x^\mu$
evaluated at these points
\begin{equation}
X^\mu_\sigma =x^\mu (\sigma).
\end{equation}
The variables to be represented as operators on the kinematical
Hilbert space are the following exponents of $X^\mu_\sigma$
\begin{equation}
U_{\sigma,k}=e^{ik_\mu X^\mu_\sigma}.
\label{vop}
\end{equation}
Here $k$ is a certain d-vector (now $d$ is the dimension of the
target space). The domain of $k$ depends on geometry of target
space. If we consider for example a string moving against
d-dimensional flat Minkowskian space-time, all the components of
$k$ may take arbitrary values. If a certain direction $x^\mu$ is
compactified say on a circle of a radius $R$ then for
$U_{\sigma,k}$ to be single-valued it is necessary that the
component $k_\mu$ take on a discrete set of values $k_\mu=n/R$
where $n$ is an arbitrary integer number.

Functions (\ref{vop}) are known in string theory as vertex operators. In our treatment they will play the same role as
Wilson loop variables do in loop quantum gravity.

To define the configuration space of the string completely we need
to impose boundary conditions on the string space-line. For closed
string the boundary condition have the following explicitly
diffeomorphism invariant form
\begin{equation}
U_{2\pi,k}=U_{0,k}.
\end{equation}
For open string the Neumann boundary condition reads
\begin{equation}
\partial_\sigma U_{\sigma,k} \Big\vert_{\sigma=0}=
 \partial_\sigma U_{\sigma,k} \Big\vert_{\sigma=2\pi}=0.
\end{equation}
This condition is invariant with respect to diffeomorphisms which
leave the end points of the string untouched.

1-forms $p_\mu d\sigma$ can be integrated over arbitrary interval of the string space-line without using any background
metric. So to each interval $[\sigma_1,\sigma_2]$ one can associate a variable
\begin{equation}
P_{\mu;\sigma_1,\sigma_2}=\int\limits_{\sigma_1}^{\sigma_2} p_\mu (\sigma ) d\sigma.
\label{psdef}
\end{equation}
These new variables have the following canonical commutation
relation
\begin{eqnarray}
\{P_{\mu ;\sigma_1,\sigma_2}, U_{\sigma,k} \}=
\left\{
\begin{array}{ccc}
i k_\mu U_{\sigma,k} & if & \sigma \in [ \sigma_1, \sigma_2 ] \\
0 & if & \sigma \not\in [ \sigma_1, \sigma_2 ]
\end{array}
\right.
\end{eqnarray}

Now one can introduce space of cylindrical functions and a scalar
product on it similarly to (\ref{cf},\ref{scpr}). The orthonormal
basis on this space has now the following form
\begin{equation}
\Psi_{\{ \sigma \}, \{ k \} }= \prod\limits_\alpha
U_{\sigma_\alpha, k^\alpha}= \prod\limits_\alpha e^{i k^\alpha_\mu
x^\mu(\sigma_\alpha) }, \label{cylfst}
\end{equation}
\begin{equation}
(\Psi_{\{ \sigma \}, \{ k \} }, \Psi_{\{ \sigma' \}, \{ k' \} })=\delta_{\{ \sigma \} \{ \sigma' \}}
\delta_{\{ k \} \{ k' \} }
\end{equation}
The diffeomorphism group of the string space-line acts on these
states  by moving points $\{ \sigma_\alpha \}$ along the string
preserving their order except for the end points of the open
string.
\begin{equation}
[{\cal U}(\phi ) \Psi_{ \{ \sigma \}, \{ k \} } ] (x^\mu)= \Psi_{
\{ \sigma \}, \{ k \} } (\phi^{-1}x^{\mu})= \Psi_{ \{ \phi \sigma
\}, \{ k \} } (x^{\mu}), \ \ \ \ \phi \in Diff
\end{equation}
In the space $H_{diff}$ of the solutions of diffeomorphism
constraint the exact positions of points $\{ \sigma \}$ are
irrelevant. Therefore a basis state in $H_{diff}$ can be
completely specified by an ordered set of d-vectors $\{ k_\mu \}$,
where $k_\mu^\alpha$ is the d-vector assigned to the point
$\sigma_\alpha$ in (\ref{cylfst}). In the case of open string
there is one to one correspondence between ordered sets of
$d$-vectors and states in $H_{diff}$. In the case of closed string
ordered sets of $d$-vectors differing from each other by cyclic
permutations correspond to the same state in $H_{diff}$.

 The scalar product on
$H_{diff}$ is defined similarly to (\ref{discpr})
\begin{equation}
\langle \{ k \} \vert \{ k' \} \rangle=
\delta_{ \{ k \} \{ k' \}}
\end{equation}

It is straightforward to calculate the spectrum of the operator
$P_{\mu ;\sigma_1,\sigma_2}$ in (\ref{psdef}). Its action is first
defined on the space $H_{aux}$. The basis states of $H_{aux}$ turn
out to be eigenstates of $P_{\mu ;\sigma_1,\sigma_2}$
\begin{equation}
P_{\mu ;\sigma_1,\sigma_2} \Psi_{ \{ \sigma \}, \{ k \} } =
\sum\limits_{\alpha \ : \ \sigma_\alpha \in [ \sigma_1, \sigma_2
]} k_\mu^\alpha \Psi_{ \{ \sigma \}, \{ k \} } .
\end{equation}
The eigenstates depend only on relative positions of points
$\sigma_\alpha$ with respect to the interval $[ \sigma_1, \sigma_2
]$ and therefore the action of the operator $P_{\mu
;\sigma_1,\sigma_2}$ on $H_{aux}$ is diffeomorphism covariant and
can be promoted  to that on the space $H_{diff}$.

\begin{figure}
\centerline{\includegraphics[width=6cm]{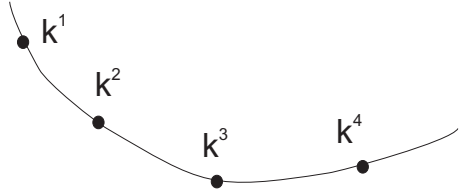}} \caption[]{A
kinematical state of the string: each ''particle'' carries a
certain fixed momentum $k$} \label{fig1}
\end{figure}

If the interval $[ \sigma_1, \sigma_2 ]$ spans the whole string
space-line, the operator $P_{\mu ;\sigma_1,\sigma_2}$ is the
operator of the total momentum of the string. Therefore a
kinematical state of the sting $\vert \{ k \} \rangle$,
$\alpha=1...N$, can be interpreted as a set of non-interacting
relativistic particles each carrying a momentum $k_\mu^\alpha$
Fig.\ref{fig1}. The momenta along compactified directions take on
discrete sets of values, while the momenta along non-compactified
directions take arbitrary values.

\section{Hamiltonian constraint}
\label{hc}

 The physical Hilbert space is the space of solutions of
both constraints diffeomorphism (\ref{difcon}) and hamiltonian
(\ref{hcon}). To construct this space one should define the
hamiltonian constraint as an operator on $H_{diff}$ and find its
kernel. It is however not straightforward to define it in an
explicitly diffeomorphism invariant way. Therefore given a state
$|\{ k \} \rangle \in H_{diff}$ the diffeomorphism invariant
hamiltonian constraint operator $\tilde{\tilde{{\cal H}}}_{diff}$
will be defined as follows
\begin{equation}
\tilde{\tilde{{\cal H}}}_{diff} \vert \{ k \} \rangle= \Pi
\tilde{\tilde{{\cal H}}}_{aux} \pi \vert \{ k \} \rangle
\label{hco}
\end{equation}
Here $\pi$ is a natural embedding $H_{diff} \rightarrow H_{aux}$,
$\tilde{\tilde{{\cal H}}}_{aux}$ is the hamiltonian constraint
operator defined on $H_{aux}$, and $\Pi$ is the projector from
$H_{aux}$ to $H_{diff}$ defined in (\ref{discpr}). The embedding
$H_{diff} \rightarrow H_{aux}$ is realized by picking up for each
$ \vert \{ k \} \rangle \in H_{diff}$ a representative $ \vert \{
\sigma \} \{ k \} \rangle \in H_{aux}$ such that $\Pi \vert \{
\sigma \} \{ k \} \rangle = \vert \{ k \} \rangle $, i.e. $\Pi \pi
= id$.
 One may note that such an  embedding  is not
unique. Moreover the resulting expression for $\tilde{\tilde{{\cal
H}}}_{diff}(N)$, the hamiltonian constraint smeared with a test
function $N$, will depend on the choice of $\pi$. It will be shown
later that the kernel of all hamiltonian constraints, smeared with
all possible test functions will nevertheless be independent of
this choice.

First we shall define the action of hamiltonian constraint
operator on $H_{aux}$. Note that $\tilde{\tilde{{\cal H}}}$ in
(\ref{hcon}) contains products of local operators. Therefore a
point splitting regularization will be required. In our treatment
the point splitting regularization will be closely related to
rewriting the constraint (\ref{hcon}) in terms of variables
$U_{\sigma ,k}$ (\ref{vop}) and $P_{\mu,\sigma_1,\sigma_2}$
(\ref{psdef}) which are well defined operators on $H_{aux}$. The
variables $U_{\sigma ,k}$ can be brought in the expression for
hamiltonian constraint (\ref{hcon}) due to the identity
\begin{equation}
k_\mu \partial_\sigma x^\mu =-ie^{-ik_\mu x^\mu }
\partial_\sigma e^{ik_\mu x^\mu}= -iU_{\sigma, -k}\partial_\sigma
U_{\sigma, k} \label{keyiden}
\end{equation}

In the present paper we consider the simplest possible geometries
of target space. Namely we take target space to be $d$-dimensional
Minkowskian space-time and allow some of $d$ dimensions to be
compactified on a torus.  In this case the metric of the target
space $g_{\mu \nu}$ can be made independent of  coordinates,
$g_{\mu \nu}=const$.  One can then introduce a  basis $\{ k_\mu^j
\}$ (not necessary orthonormal) in the target space so that the
metric on it can be written as
\begin{equation}
g_{\mu \nu}=\sum\limits_{ij} V_{ij}  k_\mu^i {k'}^j_\nu.
\label{gexp}
\end{equation}
This equation for $V_{ij}$ and $k_\mu^j$ have many different
solutions. In particular one can take
\begin{equation}
V_{ij}=\eta_{ij}  \ \ \ \  k_\mu^i={k'}_{\mu}^{i} = e_\mu^i,
\end{equation}
where $\eta_{ij}$ is the Minkowskian metric and $e_\mu^i$ is an
orthonormal basis.
%Note that for the identity (\ref{keyiden}) to be true it is
%necessary that $\partial_\sigma k^i_\mu=0$ i.e. all the components
%of the vectors of the local basis should be constant along string
%line. On the other hand for general geometry of target space the
%components of the metric $g_{\mu \nu}$ can not be made independent
%on the point. Therefore the decomposition (\ref{gexp}) should be
%made in such a way that the possible dependence of the metric
%$g_{\mu \nu}$ on the point reside in coefficients $v_{ij}$ while
%the components of the local basis $k^i_\mu$ remain constant.
%
Now by using the identity (\ref{keyiden}) hamiltonian constraint
(\ref{hcon}) can be then rewritten in the form
\begin{equation}
\tilde{\tilde{{\cal H}}}=p_\mu p^\mu - T^2\sum\limits_{ij}V_{ij}
U_{\sigma, -k^i-{k'}^j}\partial_\sigma
 U_{\sigma, k^i}\partial_\sigma U_{\sigma, {k'}^j}.
\label{hc1}
\end{equation}
The use of the identity (\ref{keyiden}) in the definition of the
hamiltonian constraint operator is similar to the construction
used in loop quantum gravity in
\cite{RovelliSmolin94,Thiemann96,Thiemann96b} where curvature
tensor at a point is replaced by a combination of the holonomies
around a set of loops originated at this point.

To promote the hamiltonian constraint in the form (\ref{hc1}) to
an operator on $H_{aux}$ one needs to regularize the operators
$p_\mu$ and $\partial_\sigma U_{\sigma, k}$ entering (\ref{hc1})
which are not well defined on the space $H_{aux}$. This can be
done via smearing them with a regularized $\delta$-function
\begin{equation}
p_\mu^\epsilon(\sigma) = \int d\sigma' \delta_\epsilon (\sigma,\sigma')p_\mu(\sigma')
\end{equation}
\begin{equation}
\partial_\sigma^\epsilon U_{\sigma, k}=
\int d\sigma' \delta_\epsilon (\sigma, \sigma') \partial_{\sigma'}
U_{{\sigma'}, k}
\end{equation}
To define a regularized $\delta$-function one should introduce a
background length on string space-line. A distance
$L(\sigma_1,\sigma_2)$ between points $\sigma_1$ and $\sigma_2$ is
defined as an integral of a positive defined weight 1 density
$l(\sigma)$ over the interval between these points:
\begin{equation}
L(\sigma_1,\sigma_2)=\int\limits_{\sigma_1}^{\sigma_2} l(\sigma) d\sigma.
\end{equation}
 In the present situation it is convenient to use
the step-like regularization of the $\delta$-function.
\begin{eqnarray}
\delta_\epsilon (\sigma, \sigma')=
\left\{
\begin{array}{ccc}
\frac{l(\sigma)}{2\epsilon} & if & L(\sigma,\sigma')<\epsilon \\
0 & if & L(\sigma,\sigma')>\epsilon
\end{array}
\right.
\end{eqnarray}
The operators $p_\mu^\epsilon$ and $\partial^\epsilon_\sigma
U_{\sigma, k}$ smeared with this $\delta$-function can be
evaluated in terms of the operators $U_{\sigma ,k}$ (\ref{vop})
and $P_{\mu,\sigma_1,\sigma_2}$ (\ref{psdef}):
\begin{equation}
p_\mu^\epsilon(\sigma)=\frac{l(\sigma)}{2\epsilon}P_{\mu,\sigma-\epsilon,\sigma+\epsilon}
\end{equation}
\begin{equation}
\partial_\sigma^\epsilon U_{\sigma,
k}=\frac{l(\sigma)}{2\epsilon}(U_{\sigma+\epsilon,k}-U_{\sigma-\epsilon,k}).
\end{equation}
As a result the regularized hamiltonian constraint (\ref{hc1}) can
also be written completely in terms these operators.
\begin{equation}
\tilde{\tilde{{\cal
H}}}^{\epsilon_1,\epsilon_2}=\frac{l^2(\sigma)}{8\epsilon_1
\epsilon_2} \Big\{ {\cal T}(\sigma,\epsilon_1,\epsilon_2)+{\cal
T}(\sigma,\epsilon_2,\epsilon_1)+ {\cal
V}(\sigma,\epsilon_1,\epsilon_2)+{\cal
V}(\sigma,\epsilon_2,\epsilon_1) \Big\} \label{h0reg}
\end{equation}
where
\begin{equation}
{\cal T}(\sigma,\epsilon_1,\epsilon_2)=
P_{\mu,\sigma-\epsilon_1,\sigma+\epsilon_1}
P_{\mu,\sigma-\epsilon_2,\sigma+\epsilon_2} \label{treg}
\end{equation}
is a ''kinetic'' term and
\begin{equation}
{\cal V} (\sigma,\epsilon_1,\epsilon_2)= T^2
\sum\limits_{ij}V_{ij} U_{\sigma, -k^i-{k'}^j}
(U_{\sigma+\epsilon_1, k^i}-U_{\sigma-\epsilon_1, k^i})
(U_{\sigma+\epsilon_2, {k'}^j}-U_{\sigma-\epsilon_1, {k'}^j})
\label{vreg}
\end{equation}
is a ''potential'' term.

Here we encounter regularization ambiguity. First, the resulting
expression for potential term (\ref{vreg}) depends on how the
parameters $\epsilon_1$ and $\epsilon_2$ relate to each other. The
expression with $\epsilon_1>\epsilon_2$ and expression with
$\epsilon_1<\epsilon_2$ will be inequivalent even at
diffeomorphism invariant level. Here the form of hamiltonian
constraint is chosen to be symmetric in $\epsilon_1$ and
$\epsilon_2$ (\ref{h0reg}).
% The advantage of this choice will be seen later.
Second, the equation (\ref{gexp}) for $k^i_\mu$ and $T^2_{ij}$ has
many different solutions. This solutions result in inequivalent
operators for potential term (\ref{vreg}).

\begin{figure}
\centerline{\includegraphics[width=10cm]{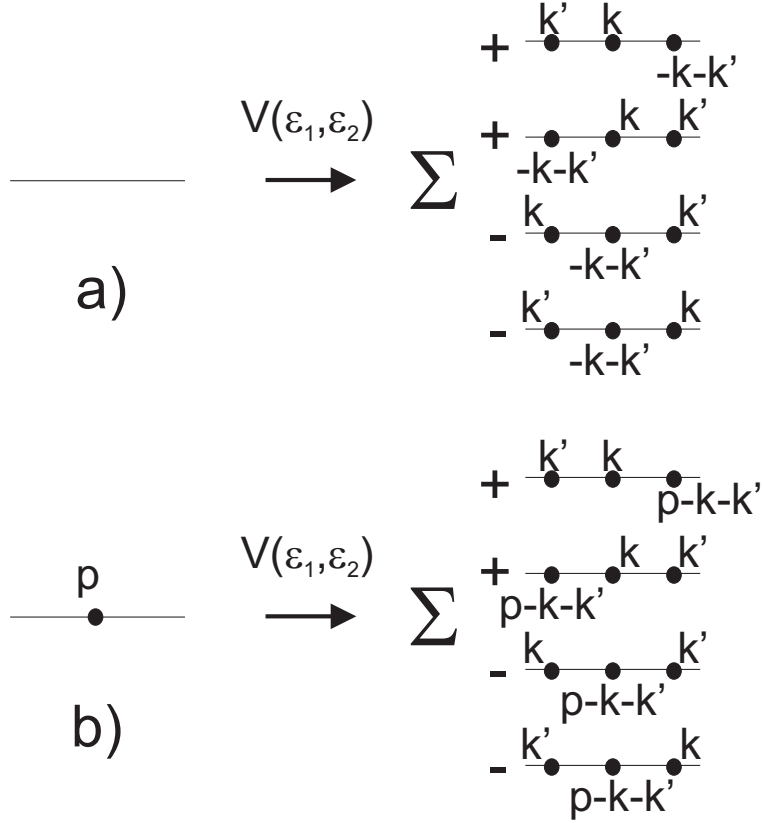}}
\caption[]{The action of the potential term ${\cal
V}(\epsilon_1,\epsilon_2)$ of the hamiltonian constraint on a
kinematical state: a) The action at an empty point b) The action
at a point with a particle having a momentum $p$} \label{fig2}
\end{figure}

Now consider how the hamiltonian constraint (\ref{h0reg}) acts on
a state $\Psi_{\{ \sigma \}, \{ k \}} \in H_{aux}$. We will begin
with considering its action at a point $\sigma_0$ which is not an
element of a set $\{ \sigma \}$ in the definition of the state
$\Psi$, i.e.  $\sigma_0 \not\in \{ \sigma \}$. Such point can be
thought of as ''empty point'', i.e. a point with no particles at
it or as a point where there is a particle with momentum equal to
zero. The action of the two kinetic terms of the hamiltonian
constraint, ${\cal T}(\sigma,\epsilon_1,\epsilon_2)$ and ${\cal
T}(\sigma,\epsilon_2,\epsilon_1)$, yields zero. Each term in the
sum in the r.h.s. of (\ref{vreg}) results in a linear combination
of four terms  adding three new particles having non-zero momentum
to the state, the total momentum of these three particles being
zero Fig.\ref{fig2}a.

When the hamiltonian constraint acts at a point where there is a
particle having a momentum $p$ the kinetic term acts as massless
Klein-Gordon operator
\begin{equation}
{\cal T}(\sigma,\epsilon_1,\epsilon_2) U_{\sigma,k}= k^\mu k_\mu
U_{\sigma,k}.
\end{equation}
The potential term (\ref{vreg}) results in a linear combination of
terms adding two new particles with non-zero momenta located in
various ways with respect to the original particle. The momentum
of the original particle is changed at the same time so that the
total momentum of three resulting particles be equal to the
original momentum of the initial particle  Fig.\ref{fig2}b.

 For neighboring points of non-zero momenta not to
affect the action of the hamiltonian constraint at the point
$\sigma_0$ $\epsilon_1$ and $\epsilon_2$ should be chosen so that
for any $\sigma_1 \in \{ \sigma \}$ $L(\sigma_0,\sigma_1)>{\rm
max}(\epsilon_1,\epsilon_2)$.This condition can not be satisfied
for fixed $\epsilon_1$ and $\epsilon_2$ as the constraint can be
applied at a point which is arbitrary close to a point with a
particles having non-zero momentum. But if $\epsilon_1$ and
$\epsilon_2$ very with $\sigma_0$ the condition
$L(\sigma_0,\sigma_1)>{\rm
max}(\epsilon_1(\sigma_0),\epsilon_2(\sigma_0))$ can be satisfied.
Of course using a position dependent regularization is something
unusual but this is the only possible regularization which is
consistent with diffeomorphism invariance. Below we will assume
that $\epsilon_1$ and $\epsilon_2$ are such functions of
$(\sigma_0)$ that the above condition is satisfied and that
$\epsilon_1 > \epsilon_2$.

Note that after applying the hamiltonian constraint at $H_{aux}$
we will then project the resulting state on $H_{diff}$. This makes
the precise positions of the tree particles added irrelevant to
the final result. Only relative order of these three points
matters. Therefore the diffeomorphism invariant result of the
action of hamiltonian constraint can be schematically shown as
''moves'' ''no particles into three particles'' ($0 \rightarrow
3$) (the action at an empty point) or ''one particle into three
particles'' ($0 \rightarrow 3$) (the action at a point with a
particle having non-zero momentum) (see Fig.\ref{fig2}).

One of the key properties of loop quantum gravity which makes it
an easily treatable theory is that the action of hamiltonian
constraint on basis states is purely combinatorial. It has
non-vanishing action only on the points where Wilson loops has
intersections while non-intersecting loops are exactly annihilated
by it \cite{JacobsonSmolin88}. It is desirable here that the
regularized hamiltonian constraint of the string have the same
property, i.e. annihilate the state while acting at a point with
no particles. There is a possibility of achieving this. One can
note that the momenta at the three new points are arbitrary, any
choice of them will reproduce the original classical expression
after removing the regulator. One therefore can make use of this
arbitrariness and  impose an additional constraint on momenta at
the three new points. The following three possible types of such
constraints result in annihilating of the empty points by the
hamiltonian constraint:
\begin{equation}
k+k'=0, \label{const1}
\end{equation}
\begin{equation}
k'=-2k, \label{const2}
\end{equation}
\begin{equation}
k=-2k' \label{const3}.
\end{equation}
This can be seen by imposing any of these constraints on linear
combination of the terms depicted at Fig.\ref{fig2}a. Then by
diffeomorphism transformation the first and the second terms in
this linear combination can be turned into the third and the
fourth and they all cancel each other. The cancellation will not
happen at a point where there is a particle with  non-zero
momentum Fig.\ref{fig2}b. Therefore there is only a finite number
of points at string space-line at which the action of the
hamiltonian constraint is non-zero.

For further constructions we will need the matrix element of the
hamiltonian constraint (\ref{h0reg}) smeared with a test function
${\cal N}$
\begin{equation}
{\cal H}[{\cal N}]=\int d\sigma {\cal N}(\sigma)
\tilde{\tilde{{\cal H}}}(\sigma) \label{smearing}
\end{equation}
between two diffeomorphism invariant states $\vert \{ k \}
\rangle$ and $\vert \{ k' \} \rangle$. Note that the operator
${\cal H} [{\cal N}]$ in (\ref{smearing}) is a non-diffeomorphism
invariant operator defined on $H_{aux}$. However given its action
on non-diffeomorphism invariant states on can define  its action
on $H_{diff}$ via (\ref{hco}).

Now one should define the smearing of the operator (\ref{h0reg})
with a test function. A consistent definition takes the following
steps:

1. Take the regularized expression (\ref{h0reg}) for
$\tilde{\tilde{{\cal H}}}^{\epsilon_1,\epsilon_2}$ acting on
$H_{aux}$ and project the result on $H_{diff}$. As it has been
mentioned above only the action at the points with a particles
having non-zero momentum will survive. Also in projecting the
result on $H_{diff}$ any dependence on regularization parameters
$\epsilon_1$ and $\epsilon_2$ will disappear.

2. Embed the obtained result into $H_{aux}$. This is necessary
because the expression (\ref{smearing}) contains a hamiltonian
constraint defined on $H_{aux}$. On the other hand we can not
simply substitute the hamiltonian constraint (\ref{h0reg}) in its
original form into the equation (\ref{smearing}) as in this case
it would contain the action on each point of the string space-line
continuum which is difficult to control. By projecting the result
on $H_{diff}$ and then embedding it back into $H_{aux}$ the action
of $\tilde{\tilde{{\cal H}}}^{\epsilon_1,\epsilon_2}$ is reduced
to that on finite number of points. The embedding of $H_{diff}$
into $H_{aux}$ is ambiguous, the positions of the points at the
string space-line can be chosen arbitrary. To restore the
information about the regularization one should place the points
added by the action of the hamiltonian constraint at the distances
$\epsilon_1$ and $\epsilon_2$ from the original point at which the
new points where added. To account the fact that the points with
particles having non-zero momenta are the only points where the
constraint have a non-zero action one should multiply the new
expression for $\tilde{\tilde{{\cal H}}}^{\epsilon_1,\epsilon_2}
(\sigma )$ by a characteristic function of $\epsilon_1$-vicinity
of those points.
\begin{eqnarray}
\tilde{\tilde{{\cal H}}}^{\epsilon_1,\epsilon_2} (\sigma )\vert \{
\sigma \}, \{ k \} \rangle = \nonumber \\
 \sum\limits_\alpha
\chi_{[\sigma_\alpha-\epsilon_1,\sigma_\alpha+\epsilon_1]}(\sigma)
\frac{l^2(\sigma)}{8\epsilon_1 \epsilon_2} \Big\{ {\cal
T}(\sigma,\epsilon_1,\epsilon_2)+{\cal
T}(\sigma,\epsilon_2,\epsilon_1)+ {\cal
V}(\sigma,\epsilon_1,\epsilon_2)+{\cal
V}(\sigma,\epsilon_2,\epsilon_1) \Big\}\vert \{ \sigma \}, \{ k \}
\rangle,
 \label{h1reg}
\end{eqnarray}
where
\begin{eqnarray}
\chi_{[\sigma_\alpha-\epsilon_1,\sigma_\alpha+\epsilon_1]}(\sigma)
= \left\{
\begin{array}{ccc}
1 & if & L(\sigma,\sigma_\alpha)<\epsilon_1 \\
0 & if & L(\sigma,\sigma_\alpha)>\epsilon_1
\end{array}
\right.
\end{eqnarray}
The multiplication by of the  expression for the hamiltonian
constraint by $\sum\limits_\alpha
\chi_{[\sigma_\alpha-\epsilon_1,\sigma_\alpha+\epsilon_1]}(\sigma)$
is the summary result of its projection on $H_{diff}$ and then
embedding into $H_{aux}$.

 3. Smear the hamiltonian constraint obtained in the previous
step with a test function  ${\cal N}$.  Because the hamiltonian
constraint is a density of the weight 2 with respect to
diffeomorphisms of the space line of the string the test function
${\cal N}$ in (\ref{smearing}) has to be a density of the weight
-1. We will need also scalar test function $N$ related to ${\cal
N}$ by
\begin{equation}
N=\frac{l(\sigma)}{2 \epsilon_2} {\cal N} \label{nscal}
\end{equation}
The relation between $N$ and ${\cal N}$ depends on a background
metric $l$, but because the test function is arbitrary this
doesn't introduce any ambiguity. By taking into account that
\begin{equation}
\chi_{[\sigma_\alpha-\epsilon_1,\sigma_\alpha+\epsilon_1]}(\sigma)
\frac{l(\sigma)}{2\epsilon_1}=\delta_{\epsilon_1}(\sigma,
\sigma_\alpha)
\end{equation}
the resulting expression can be written in the form:
\begin{eqnarray}
{\cal H}[{\cal N}]\vert \{ \sigma \}, \{ k \} \rangle= \nonumber
\\
\int d\sigma N(\sigma) \sum\limits_\alpha
\delta_{\epsilon_1}(\sigma, \sigma_\alpha) \frac{1}{2} \Big\{
{\cal T}(\sigma,\epsilon_1,\epsilon_2)+{\cal
T}(\sigma,\epsilon_2,\epsilon_1)+ {\cal
V}(\sigma,\epsilon_1,\epsilon_2)+{\cal
V}(\sigma,\epsilon_2,\epsilon_1) \Big\}\vert \{ \sigma \}, \{ k \}
\rangle. \label{hsmr}
\end{eqnarray}

4. Finally we can remove the regulator in the equation
(\ref{hsmr}) and perform the integration over $\sigma$. The result
will be
\begin{equation}
{\cal H}[ {\cal N} ] \vert \{ \sigma \}, \{ k \} \rangle
=\sum\limits_{ \{ \sigma \} } N(\sigma_\alpha) {\cal
H}(\sigma_\alpha) \vert \{ \sigma \}, \{ k \} \rangle,
\end{equation}
where
\begin{equation}
{\cal H}(\sigma_\alpha)= \lim_{\epsilon_1,\epsilon_2 \rightarrow
0}\Big\{ {\cal T}(\sigma,\epsilon_1,\epsilon_2)+{\cal
T}(\sigma,\epsilon_2,\epsilon_1)+ {\cal
V}(\sigma,\epsilon_1,\epsilon_2)+{\cal
V}(\sigma,\epsilon_2,\epsilon_1) \Big\}
\end{equation}

By definition of scalar product on diffeomorphism invariant states
(\ref{discpr}) one can write the matrix element of $H[{\cal N}]$
as follows
\begin{eqnarray}
\langle \{ k' \} \vert H [{\cal N}] \vert \{ k \} \rangle=
\sum\limits_{ \{ \sigma \} } N(\sigma_\alpha) \langle \{ k' \}
\vert H(\sigma_\alpha) \vert \{ \sigma \}, \{ k \} \rangle
\label{me1}
\end{eqnarray}
One can see that the resulting expression for matrix element is
not diffeomorphism invariant. It depends on the locations of the
points $\sigma_\alpha$ which represent the embedding $\pi$ of a
state $\vert \{ k \} \rangle \in H_{diff}$ to a state $\vert \{
\sigma \}, \{ k \} \rangle \in H_{aux}$. However to obtain a
physical matrix element a functional integration over $N(\sigma)$
should be performed and as it was shown in \cite{Rovelli98} the
dependence on the embedding $\pi$ can be removed from the
resulting expression.

A matrix element entering the sum in the r.h.s. of (\ref{me1}) can
be evaluated explicitly by using the expression (\ref{h0reg}) for
the regularized  hamiltonian constraint
\begin{eqnarray}
\langle \{ k' \} \vert H(\sigma_\alpha) \vert \{ \sigma \}, \{ k
\} \rangle &=& k_{\alpha,\mu} k_{\alpha}^\mu \langle \{ k' \}
\vert  \{ k \} \rangle-
 T^2\sum\limits_{ij} V_{ij} \langle \{ k' \}
\vert K_{\alpha,ij} \{ k \} \rangle
\nonumber \\
&=& k_{\alpha,\mu} k_{\alpha}^\mu \delta_{ \{ k' \}   \{ k \} }-
T^2\sum\limits_{ij} V_{ij} \delta_{ \{ k' \}  K_{\alpha,ij} \{ k
\} }
\end{eqnarray}
Here the operator $K_{\alpha,ij}$ is defined as follows (recall
that the diffeomorphism invariant state $ \vert \{ k \} \rangle$
is defined as an ordered set of momenta $(k_1,...,k_N)$):
\begin{eqnarray}
K_{\alpha,ij}(k_1,...,k_\alpha,...,k_N)=
(k_1,...,k_i,k'_j,k_\alpha,...,k_N)
+(k_1,...,k_\alpha,k'_j,k_i,...,k_N)
\nonumber \\
-(k_1,...,k'_i,k_\alpha,k_j,...,k_N)
-(k_1,...,k'_j,k_\alpha,k_i...,k_N) + (k_i \leftrightarrow k'_j)
\label{vdef1}
\end{eqnarray}

In order to allow the this hamiltonian constraint to convey
momentum from one point in original state to another one should
take the hamiltonian constraint in  (\ref{spf}) to be symmetric
one ${\cal H}[N]={\cal H}_{sym}[N]$, i.e. to allow it not only
create but also annihilate new points in a state,
\begin{equation}
\langle \{ k' \} \vert  {\cal H}_{sym}[N] \vert \{ k \} \rangle=
\langle \{ k' \} \vert  {\cal H}[N] \vert \{ k \} \rangle+
\overline{ \langle \{ k \} \vert  {\cal H}[N] \vert \{ k' \}
\rangle }
\end{equation}

Now given the expression for the hamiltonian constraint it's
straightforward to calculate the quantum constraint algebra. It
turns out to be very similar to that in loop quantum gravity
\cite{LewandowskiMarolf},\cite{LewandowskiEtAl}:
\begin{eqnarray}
\big[ {\cal D},{\cal D} \big] \sim {\cal D}, \nonumber \\
\big[ {\cal H},{\cal D} \big] \sim {\cal H}, \nonumber \\
\big[ {\cal H},{\cal H} \big] =0
\end{eqnarray}
Because of the last equation this algebra is not isomorphic to the
classical constraint algebra. This is the first problem we
encounter.

Now one can try to solve the constraint equation defined. First of
all one can note that the vacuum (the state with no particles) is
a solution of both constraints. The first non-trivial solution is
a state with a single particle located at a closed string:
\begin{equation}
\vert \{ k \} \rangle =e^{ik_{1,\mu}X^\mu(\sigma_1)}.
\label{tacsol}
\end{equation}
It is easy to see that the four terms resulting from the potential
term of the hamiltonian constraint and depicted at Fig.\ref{fig2}b
cancel each other if we connect the free ends of the string
together thereby forming a closed string. The kinetic term in the
hamiltonian constraint disappears provided that the momentum
$k_{1,\mu}$ entering (\ref{tacsol}) is null,
$k_{1,\mu}k_{1}^{\mu}=0$. This state should be identified with the
well known tachionic mode of the string. This is because the
particle is a scalar and also in string theory the tachionic state
is produced by the insertion of vertex operator of the form
(\ref{tacsol}) into string world-sheet. Generally due to the
anomaly in string theory the states of the string should be taken
to be solutions of a shifted hamiltonian constraint ${\cal
H}'={\cal H} + constant$. The constant would play the role of the
squared mass of the mode and being negative would make the mode
tachionic. However the present regularization of the hamiltonian
constraint produces an anomaly-free constraint algebra and
therefore the scalar string mode have to be massless. Another
problem is that the solution (\ref{tacsol}) exists only for the
closed string, this can not be generalized to the case of the open
string.

The following non-trivial solutions to all the constraints are
expected to be higher modes of the string. Because we don't have
the higher modes vertex operators at our disposal they are to be
constructed from tachionic vertex operators. The solution to all
the constraints can be sought in a form of an infinite linear
combination of terms with various numbers of tachionic vertex
operators inserted. This can be derived via an expansion of the
projector on the physical states. This is what the next section is
devoted to.

\section{The projector on the physical states}
The construction of projector on the physical Hilbert space  of
the string and its interpretation will closely followed to those
introduced in \cite{Rovelli98}. It is based on perturbative
expansion of the functional integral
\begin{equation}
P \sim \prod\limits_\sigma \delta ({\cal H}(\sigma)) \sim \int
[DN] e^{-i\int d\sigma N(\sigma)  {\cal H}(\sigma)}
\end{equation}
Given the matrix element of this operator between general two
general states from $H_{diff}$
\begin{equation}
\langle \{ k' \} \vert P \vert \{ k \} \rangle = \int [DN] \langle
\{ k' \} \vert e^{-iH[N]} \vert \{ k \} \rangle \label{fi}
\end{equation}
we can define the physical Hilbert space $H_{phys}$ over the
pre-Hilbert space $H_{diff}$ by the quadratic form
\begin{equation}
\langle \{ k' \} \vert  \vert \{ k \} \rangle_{phys}= \langle \{
k' \} \vert P \vert \{ k \} \rangle.
\end{equation}

\begin{figure}
\centerline{\includegraphics[width=10cm]{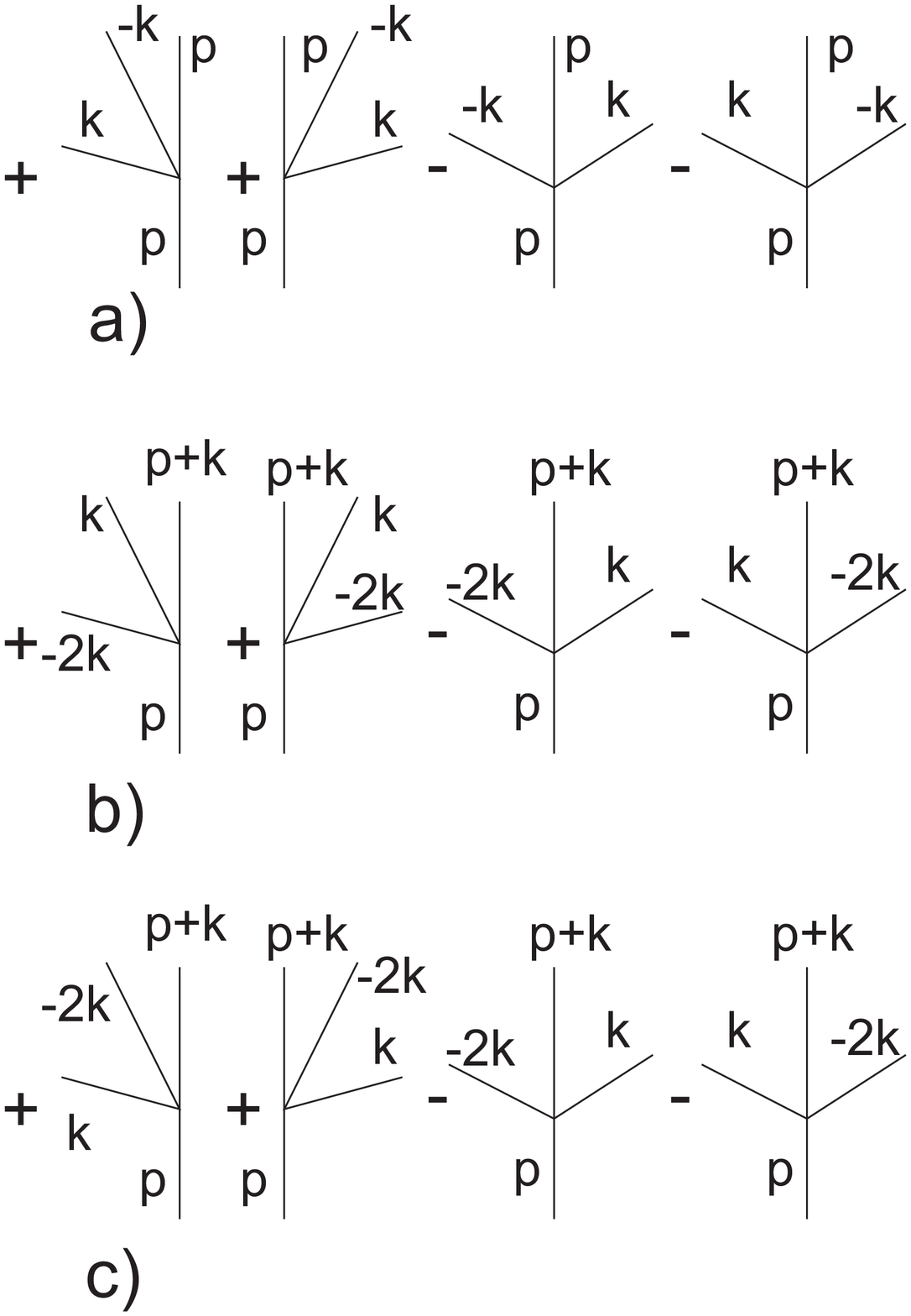}}
\caption[]{Sets of diagrams appearing as a result of the action of
the hamiltonian constraint at a point with a particle having a
momentum $p$ ; a) with constraint (\ref{const1}), b) with
constraint (\ref{const2})}, c) with constraint (\ref{const3})
\label{fig3}
\end{figure}

In \cite{Rovelli98} the functional integral is regularized by
\begin{equation}
\vert N(\sigma) \vert < N_{max} , \label{creg}
\end{equation}
where $N_{max}$ is a certain constant. The physical limit is
recovered for $N_{max} \rightarrow \infty$. Then the exponent in
(\ref{fi}) is replaced by6 its Taylor series. The resulting
expression is therefore involve matrix elements of different
powers of ${\cal H}[N]$ which as we have seen in the previous
section depend on the values of $N$ at a certain set of points and
therefore are not diffeomorphism invariant. However the expression
(\ref{fi}) as well as the regularization (\ref{creg}) are
diffeomorphism invariant. Therefore we can insert an integration
over the diffeomorphism group in the expression (\ref{fi}). The
resulting expression then reads
\begin{eqnarray}
\langle \{ k' \} \vert P_{N_{max}} \vert \{ k \} \rangle=
\int_{Diff}D\phi \int_{| N(\sigma) |< N_{max}} DN \langle {\cal
U}(\phi) \{ k' \} \{ \sigma' \} \vert \left(
\sum\limits_{n=0}^{\infty} \frac{(-i)^n}{n !} ({\cal H}[N])^n
\right)
 \vert \{ k \} \rangle.
\label{spf}
\end{eqnarray}
It can be interpreted as a sum over planar Feynmann diagrams with
vertices depicted on Fig.\ref{fig3}.

Given the hamiltonian constraint found in the previous section,
the spin-foam diagrams of this model become very similar to those
of two-dimensional $\phi^4$-model. They are comprised by
four-valent vertices and lines connected them. There is however a
certain difference between them. First, no integration over the
momenta in closed loops is required. The momentum flowing around a
closed loop is a free parameter which appears in regularization.
$\phi^4$-model can however be recovered by taking a sum over all
the possible regularization parameters. Another feature of
$\phi^4$ which is lacking in the present model is the possibility
of $2 \rightarrow 2$ move. The later probably reflects the general
difficulty with application of canonical quantization to a
generally covariant theory. Space-time covariance get lost in
canonical approach and to recover it a resulting theory should be
modified e.g. by adding some extra moves
\cite{MarkopoulouSmolin97}, \cite{ReisenbergerRovelli}. Finally,
an additional constraint (\ref{const1}) or (\ref{const2}) or
(\ref{const3}) on momenta meeting at a vertices needs to be
imposed. To recover $\phi^4$-model one should make the resulting
theory free of this constraints. One can easily see that the
constraints (\ref{const2}) and  (\ref{const3}) are not invariant
with respect to renormalization of the tension which in the
present case plays the role of coupling constant. By applying
block transformation to four valent vertices subject to these
constraint one can recover the general vertex of $\phi^4$-model
with the only constraint being the conservation of momentum. The
vertices with constraint (\ref{const1}) remain unchanged under
renormalization, but this is not an interesting case because such
vertices do not convey momenta from one point to another.
Therefore, one can conclude that in any case we should end up with
the two-dimensional $\phi^4$-model. The only freedom remained is
the domain of integration over the momenta flowing around closed
loops.

\section{Hamiltonian constraint, second version}
\label{hc2}

 The suggested version of the hamiltonian constraint
leave us with a theory which is considerably different from
ordinary string theory. There are several indications that string
theory in its usual form can probably not be recovered from the
model obtained. Firstly, in conformal field theory the expectation
value of a product of three vertex operators is not equal to zero.
On the other hand the model we obtained is similar to
$\phi^4$-model which does not contain a transition involving three
vertex operators. Nor can three valent vertices be recovered via
block transformation. Besides, the model for representing a
discretized string is generally taken to be $\phi^3$
\cite{Ambjorn}. Another difference between them is that as it was
mentioned in section \ref{hc} the first version of hamiltonian
constraint is anomaly-free and the same is true of the
diffeomorphism constraint. Alternatively, in ordinary string
theory quantum constraint algebra contains an anomaly which is
absent only in critical dimension of target space $d=26$.

\begin{figure}
\centerline{\includegraphics[width=10cm]{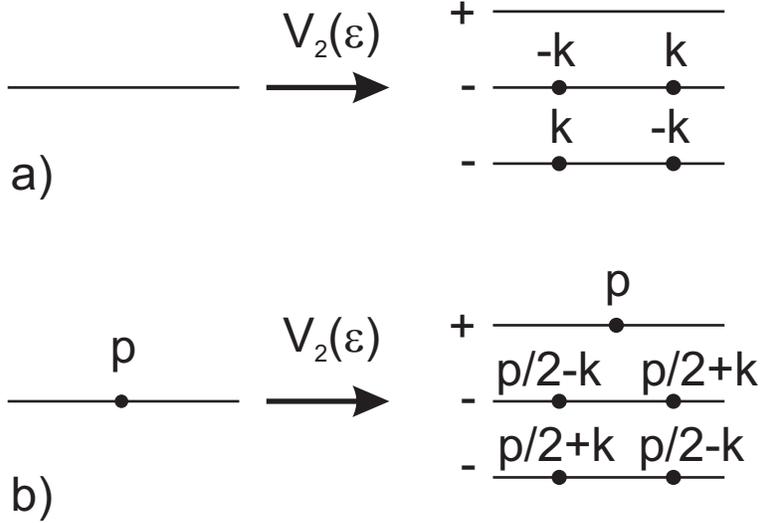}}
\caption[]{Action of the potential term of the hamiltonian
constraint defined in (\ref{vreg2}) on a kinematical state. a) at
an empty point. b) at a point with a particle having a nonzero
momentum} \label{fig4}
\end{figure}

There is however another regularization of the hamiltonian
constraint which results in a model having much more similarity
with ordinary quantization of the string. By making use of
regularization ambiguity one can get a theory which resemble
$\phi^3$-model and contains an anomaly which can be cancelled by
tuning the parameters of the theory. Given most general
regularization of ${\cal H}$ (\ref{h0reg}) one can first set
$\epsilon_1=\epsilon_2=\epsilon$, momenta at coinciding points
being added to each other. Then one can require that the action of
the hamiltonian constraint cancel the momenta at the points where
it acts , i.e. $p(\sigma)-k-k'=0$. This is another piece of
position dependence in the regularization (recall that $\epsilon$
is also position dependent). The resulting hamiltonian constraint
can be written as follows,
\begin{equation}
\tilde{\tilde{{\cal H}_2}}^{\epsilon}=\frac{l^2(\sigma)}{4
\epsilon^2} \Big\{ {\cal T}_2(\sigma,\epsilon)+ {\cal
V}_2(\sigma,\epsilon) \Big\} \label{h0reg2}
\end{equation}
where
\begin{equation}
{\cal T}_2(\sigma,\epsilon)=
P_{\mu,\sigma-\epsilon,\sigma+\epsilon}
P_{\mu,\sigma-\epsilon,\sigma+\epsilon} \label{treg2}
\end{equation}
 and
\begin{eqnarray}
&&{\cal V}_2 (\sigma,\epsilon) \vert \{ \sigma \}, \{ p \} \rangle
\nonumber \\
 &=& T^2\sum\limits_{ij}V_{ij} U_{\sigma, -p}
(U_{\sigma+\epsilon, p/2-k^i}-U_{\sigma-\epsilon, p/2-k^i})
(U_{\sigma+\epsilon, p/2+k^i}-U_{\sigma-\epsilon, p/2+k^i})\vert
\{ \sigma \}, \{ p \} \rangle
\nonumber \\
&=&T^2\sum\limits_{ij}V_{ij}[2U_{\sigma, -p}U_{\sigma+\epsilon,
p}- U_{\sigma+\epsilon, p/2+k^i}U_{\sigma-\epsilon,
p/2-k^i}-U_{\sigma+\epsilon, p/2-k^i}U_{\sigma-\epsilon, p/2+k^i}]
\vert \{ \sigma \}, \{ p \} \rangle
\nonumber \\
&-&T^2\sum\limits_{ij}V_{ij}[U_{\sigma, -p}U_{\sigma+\epsilon,
p}-U_{\sigma, -p}U_{\sigma-\epsilon, p} ] \vert \{ \sigma \}, \{ p
\} \rangle
 . \label{vreg2}
\end{eqnarray}
In the above equation by $p$ we meant $p(\sigma)$ -- the momentum
at the point of the string where the operator acts. It is easy to
see that as $\epsilon \rightarrow 0$ $U_{\sigma,
-p}U_{\sigma+\epsilon, p}$ is the identity operator and
$[U_{\sigma, -p}U_{\sigma+\epsilon, p}-U_{\sigma,
-p}U_{\sigma-\epsilon, p} ]/\epsilon$ is a diffeomorphism. The
identity operator will introduce a constant term into the
hamiltonian constraint, which is similar to the appearance of a
central term in  the Virasoro algebra in string theory in the Fock
representation from commutator between the creation and
annihilation operators. Thus, this version of the hamiltonian
constraint contains moves $0\rightarrow 0$, $0\rightarrow 2$, $1
\rightarrow 1$, and $1 \rightarrow 2$, Fig.\ref{fig4}.
Moves $0 \rightarrow 0$ and $1 \rightarrow 1$ correspond to an
additive constant $c=\sum\limits_{i,j=1}^d T_{ij}$ in the
hamiltonian constraint. This term results in an anomaly in the
constraint algebra as the commutator between the hamiltonian and
the diffeomorphism constraints should close on the hamiltonian
constraint but the additive constant commutes with constraint and
therefore the hamiltonian constraint will not be reproduced. It is
natural to suggest that this is analogous to the well known
conformal anomaly in string theory, $c$ playing the role of the
central charge. In string theory conformal anomaly can be
cancelled by tuning the dimension of space, $d=26$. In our
treatment $c$ can be thought of as a vacuum energy of the string
and (when the hamiltonian constraint acts at a point with no
particles) or when the hamiltonian constraint acts at a point
where there is a particle it can be interpreted as a squared mass
of this particle (in this case the massless Klein-Gordon  operator
$k^\mu k_\mu$ acting on this particle is replaced by $k^\mu k_\mu
+ c$).

A natural question then arises: is there a possibility to cancel
this anomaly by tuning parameters of the theory? If so, will it
constraint the dimension of space to be equal to 26? The answer on
this question would require a more detailed consideration. For the
present, one can give an argument that the central charge can be
changed by changing the parameters of the theory. As in ordinary
interacting quantum field theory, the transition amplitude between
two states can be represented as a sum over Feynmann diagrams. The
contributions from some of these diagrams can be taken into
account by a redefinition of the constants of the theory. The same
can be done here. First, the moves $0 \rightarrow 2$ give rise to
vacuum loops in Feynmann diagrams. All these diagrams contribute
to the vacuum energy of the theory which was initially equal to
$c$.
\begin{figure}
\centerline{\includegraphics[width=10cm]{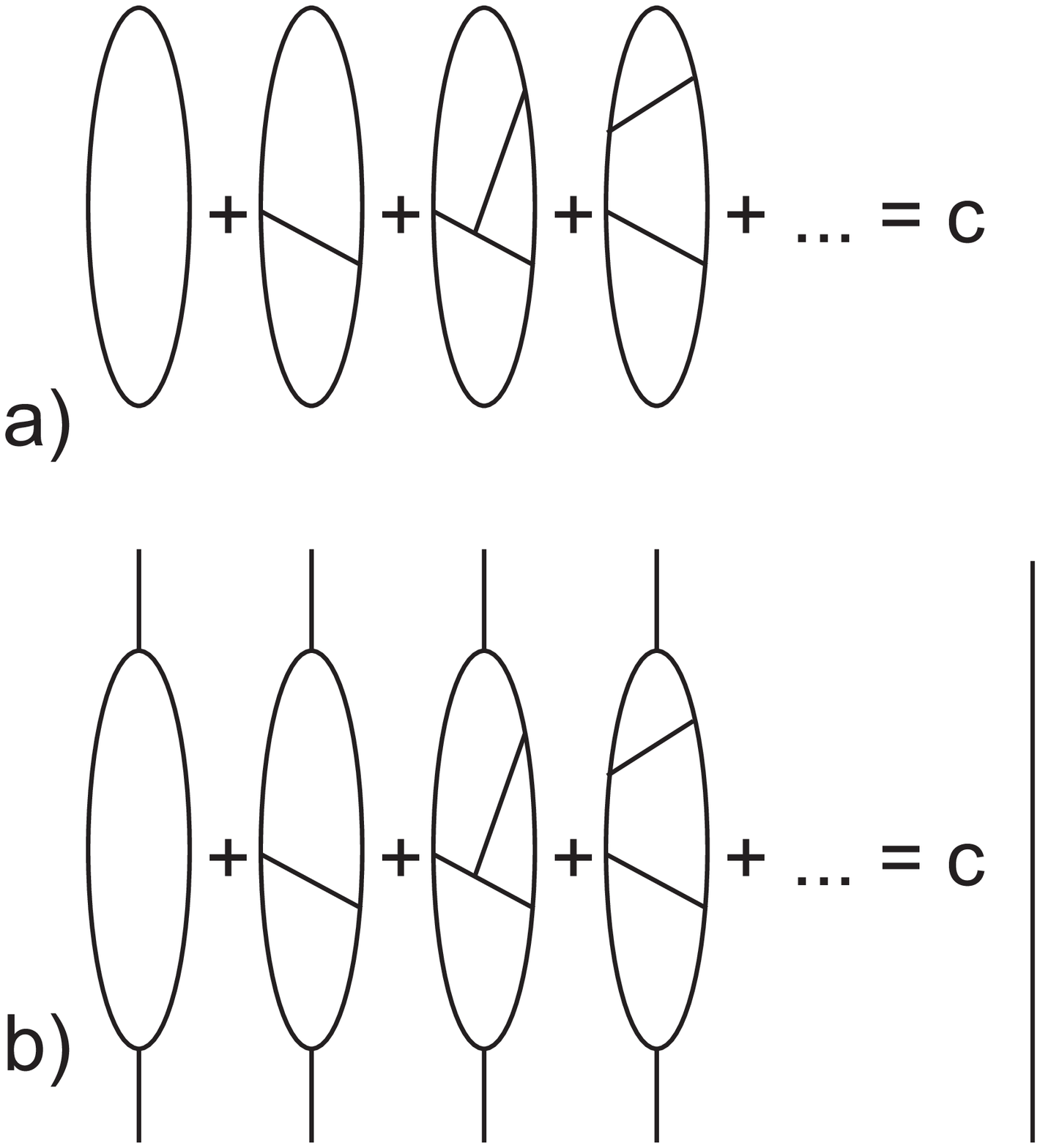}} \caption[]{
a) corrections to the vacuum energy of the string;  b) corrections
to the mass of the particles} \label{fig5}
\end{figure}
Therefore if the sum over all vacuum loops is equal to $c$ as it
is shown on Fig.\ref{fig5}a, the vacuum energy of the string is
equal to zero. Similarly, diagrams shown on Fig.\ref{fig5}b
contribute to the mass of the particle and can cancel it under
certain conditions. It is important to note that the condition of
setting the vacuum energy to zero is consistent with the condition
of setting the masses of the particles to zero. This is due to the
apparent one to one correspondence between the set of diagrams
shown on Fig.\ref{fig5}a and that shown on Fig.\ref{fig5}b. This
consistency allow one in principle to cancel the anomaly by
setting the vacuum energy and masses of the particles to zero.

\section{Discussion}
The main issue the proposed model is intended to address is the
problem of continuum limit in loop quantum gravity. Ordinary
string theory based on Fock space quantization have the right
continuum limit. Therefore in the present context this problem
reduces to that of finding a relation between the resulting model
we obtained and the ordinary quantum string.

There is another indication that the right way of looking for
continuum limit of this model is through the relation between it
and ordinary string theory. Let us recall that the problem of
continuum limit in loop quantum gravity consists of two parts.
First part is the problem of the existence of moves which would
convey interaction between neighboring vertices of spin network.
This problem has no analog in the present model: both versions of
the hamiltonian constraint proposed allow a creation of a particle
by one particle in an initial state and its subsequent absorbtion
by a neighboring particle. If the first part of the problem is
solved there appears another one: in continuum limit perturbations
of a state of the theory should propagate arbitrary far, while in
a general discrete model with interaction between nearest
neighbors perturbations decay at a certain distance. The later can
be interpreted as that the particles which are responsible for
mediating the interaction acquire a mass. This problem is
generally solved by tuning parameters of a model so that the mass
of these particles is set to zero. It is interesting to note in
this connection that by requiring that the resulting theory be
anomaly-free we set the masses of the particles associated with
vertex operators to zero. However it is still unclear whether
these particles can be identified with those responsible for
mediating perturbations of a state of the string.

The easiest way to establish the equivalence between this model
and ordinary string theory would be to show that the partition
function of three vertex operators, which is a well known quantity
in string theory, coincides with the amplitude of the move $1
\rightarrow 2$ given by the hamiltonian constraint with
renormalized parameters. This is not straightforward to do because
to find the renormalized values of parameters one should perform a
summation over all the diagrams in the series. Therefore, it will
be necessary to find a natural small parameter in which the
expansion is made.

In QFT models such as $\phi^3$ the only such parameter is the
coupling constant. In our treatment the role of coupling constant
is played by string tension. The model in the limit $T \rightarrow
0$ corresponds to kinematical picture, the only difference being
that the particles should now obey Klein-Gordon equation $k_\mu
k^\mu =0$. The hamiltonian constraint of the model can therefore
be divided into two parts: a free hamiltonian and an interaction
term. One can then construct ''interaction representation'' for
this model by associating an exact solution of the free
hamiltonian to each line in the diagrams and an interaction term
to each vertex. In this case Feynmann diagrams can be understood
as an expansion in powers of string tension. Therefore this model
should work well in the limit of small tension.

It is interesting to consider the role played by T-duality in this
model. The critical radius of compactification below which string
modes become redundant is proportional to the inverse tension of
the string. The kinematics of this model corresponds to zero
tension and therefore the critical radius of compactification is
equal to infinity. This means that we are always below the
critical radius and there is a certain redundancy in the
description of kinematical state. This redundancy manifest itself
in the fact that e.g. a state with two particles having momenta
$k^1$ and $k^2$ is kinematically equivalent with a state with one
particle having a momentum $k^1+k^2$. Therefore all the
kinematical states of the string can be made of ''elementary''
particles having the smallest possible momentum. This picture has
a close similarity with string bit models
\cite{SusskindKlebanov88,Thorn93}. How to extend this picture on
dynamics is still unclear.

String bit models describe string theory in light cone frame
$x^-=x^0-x^d=0$. In this connection it is interesting to note that
in light cone frame the considerations of this paper would undergo
a considerable simplification. Recall that there was a problem
with making  the hamiltonian constraint ``combinatorial'', i.e.
acting only on a finite set of points. In light cone frame this is
satisfied automatically and therefore neither an additional
constraint on the regularization parameters (like in section
\ref{hc}) nor renormalization of the vacuum energy (like in
section \ref{hc2}) is required. This is because in this frame the
longitudinal momentum is always positive $p_-\geq 0$. This
property together with conservation of momenta $\sum p_-^{i}=0$
results in setting all the longitudinal momenta created by an
action of hamiltonian constraint to zero $p_-^i-=0$. In quantum
field theory in light cone frame the point $p_-=0$ is called
``zero mode''. It is a singular point of the theory and a
regularization is required here. One generally introduces a
regularization condition $p_->\epsilon>0$ and therefore zero mode
can be neglected. In the present case there is a singularity in
the hamiltonian constraint for small added momentum $k$. It can be
seen from the equation (\ref{gexp}) that as $k_\mu^i \rightarrow
0$, $T^2_{ij} \rightarrow \infty$. Therefore the infrared cutoff
will be necessary in this theory anyway  and this in particular
should rule the zero mode out.

\section*{Acknowledgements}
I am grateful first of all to Lee Smolin for encouragement and
numerous suggestions during the course of work. Conversations with
Kirill Krasnov and Don Marolf where also very helpful. Also I
would like to thank Valentin A. Franke and Sergei Paston for
discussions on this work at early stage.

\end{document}